\documentclass[12pt,preprint]{aastex}

\def\ms{m~s$^{-1}$}
\def\ks{km~s$^{-1}$}
\def\msini{$M_P\sin{i}$}

\def\msun{$M_{\odot}$}
\def\mjup{$M_{\rm Jup}$}
\def\rsun{$R_{\odot}$}
\def\chisq{$\sqrt{\chi^2_\nu}$}
\def\plmn{~$\pm$~}
\def\star{HD\,185269}
\def\rms{10.1}
\def\chisq{1.14}
\def\vmag{6.67}
\def\bv{0.606}
\def\mv{3.29}
\def\dist{47}
\def\sptype{G0IV}
\def\mstar{1.28}
\def\rstar{1.88}
\def\fe{+0.11}
\def\vsini{6.1}
\def\teff{5980}
\def\logg{3.94}
\def\rhk{-5.14}
\def\sval{0.14}
\def\age{4.2}
\def\prot{23}

\def\nobseff{30}
\def\jitter{5.0}
\def\perr{0.001}
\def\period{6.838}
\def\terr{0.18}
\def\tp{2453154.1}
\def\ttr{2453153.1}
\def\ttrerr{0.16}
\def\eerr{0.04}
\def\ecc{0.30}
\def\oerr{6.8}
\def\om{173}
\def\kerr{4.5}
\def\kamp{91}
\def\msini{0.94}
\def\a{0.077}

\begin{document}
\title{An Eccentric Hot Jupiter Orbiting the Subgiant\ \star$^1$}

\author{ John Asher Johnson\altaffilmark{2}, 
  Geoffrey W. Marcy\altaffilmark{2},
  Debra	A. Fischer\altaffilmark{3}, 
  Gregory W. Henry\altaffilmark{4}, 
  Jason T. Wright\altaffilmark{2},
  Howard Isaacson\altaffilmark{3}, 
  Chris McCarthy\altaffilmark{3}}

\email{johnjohn@astron.berkeley.edu}

\altaffiltext{1}{Based on observations obtained at the Lick
  Observatory, which is operated by the University of California}
\altaffiltext{2}{Department of Astronomy, University of California,
Mail Code 3411, Berkeley, CA 94720}
\altaffiltext{3}{Department of Physics \& Astronomy, San Francisco
  State University, San Francisco, CA 94132}
\altaffiltext{4}{Center of Excellence in Information Systems, Tennessee
  State University, 3500 John A. Merritt Blvd., Box 9501, Nashville, TN 37209}

\begin{abstract}
We report the detection of a Jupiter--mass planet in a 
\period~d orbit around the \mstar~\msun\ subgiant \star. The 
eccentricity of \star b ($e = \ecc$) is unusually large compared to
other planets within 0.1~AU of their stars. 
Photometric observations 
demonstrate that the star is constant to $\pm$0.0001 mag on the radial
velocity period, strengthening our interpretation of a planetary companion.
This planet was detected as part of our
radial velocity survey of evolved stars located on the subgiant branch
of the H--R diagram---also known as the Hertzsprung Gap. 
These stars, which have masses between 1.2 and 2.5~\msun, play an
important role in the investigation of the frequency of extrasolar
planets as a function of stellar mass.
\end{abstract}


\section{Introduction}

The assemblage of $\sim 185$ known extrasolar planets\footnote{For the
  updated catalog of extrasolar planet and their parameters see
  http://exoplanets.org.} 
has revealed important relationships between the physical
characteristics of stars and the likelihood that they harbor 
planets \citep{udry03, butler06}. One example is  
the strong correlation between planet occurrence and stellar metallicity 
\citep{gonzalez97, santos04, fischer05b}. The planet--metallicity
relationship can be 
understood in the context of the core accretion model of planet
formation, in which Jovian planets begin as large rocky cores 
that then accrete gas from their surrounding protoplanetary disks once they
exceed a critical core mass \citep[e.g.][]{pollack96}. The growth of
these embryonic cores is enhanced by increasing the surface density of 
solid particles in the disk, which is related to the metallicity of the
star/disk system \citep{alibert05, ida05a}.  

The growth rate of rocky cores in protoplanetary disks is also related
to the total disk mass. Assuming that the disk mass increases with the
mass of the central star, the planet occurrence rate should therefore 
also correlate with stellar mass. \citet{laughlin04} showed that the 
lower surface densities of M~dwarf protoplanetary disks impede
the growth of Jupiter--mass planets. The relationship between 
stellar mass and planet occurrence was studied in further
detail by \citet{ida05b} for a larger range of stellar masses. Based
on their Monte Carlo simulations, they predict a positive
correlation between the number of detectable planets and stellar mass
up to about 1~\msun. However, there is some debate about whether
surface density of solid material in protoplanetary disks is
proportional to the mass of the central star. By assuming that the
initial conditions of the disk are independent from the mass of
star, \citet{kornet06} find that the occurrence rate of planets is
inversely related to stellar mass. 

The prediction that low--mass stars should have a lower frequency of
Jovian planets is in accordance with observations; only one M~dwarf,
GL~876, is known to harbor Jupiter--mass planets \citep{marcy01}. From
their survey of 90 M~dwarfs, \citet{endl06} estimate that less than
1.27\% of such low--mass stars harbor
Jovian--mass planets within 1~AU. \citet{laws03} studied the planet
occurrence rate in light of the current observational data using the
larger set of FGK stars surveyed 
as part of the California \& Carnegie Planet Search (CCPS, the full,
updated target list can be found in
\citet{wright04b})). They found evidence that the planet 
rate decreases for lower stellar masses and peaks near 1.0~\msun. However,  
the planet rate is poorly constrained for stellar masses greater than
about 1.2~\msun\ due to the small number of intermediate--mass stars
($1.3 \lesssim M_{star} \lesssim 3$~\msun) in the CCPS sample. This
dearth of massive stars is due to an 
observational bias since main--sequence stars with spectral 
types earlier than F8 tend to be fast rotators \citep{donascimento03},
have fewer spectral lines, and display a large 
amount of  chromospheric activity \citep{saar98}. These features 
result in a decrease in the radial velocity precision attainable from the
spectra of main--sequence stars more massive than $\sim
1.2$~\msun \citep{galland05}. Thus, early--type dwarfs are not
typically monitored as part of most radial velocity surveys. 

One method to circumvent these difficulties is to observe
intermediate--mass stars after they evolve into the region of the H--R
diagram between the main--sequence and red giant 
branch---also known as the Hertzsprung Gap (HG). 
After stars have expended their core hydrogen fuel sources their
radii expand, their photospheres cool, and convection sets in 
below their photospheres. Convective motion and stellar
rotation drive magnetic fields that couple with an expanding stellar
wind and act as a rotational brake \citep{gray85, schrijver93,
 donascimento00}. The cooler atmospheres and slower rotational
velocities of evolved stars lead to an increased number of narrow
absorption lines in their spectra, making HG stars better suited for
precise radial velocity measurements than their main--sequence progenitors.  

We are conducting a radial velocity survey of 159 HG stars to
search for planets orbiting intermediate--mass stars. We describe the selection
criteria of our sample of stars in \S~\ref{sample}. We present here 
the first planet detection from our sample of HG 
stars: an eccentric hot jupiter orbiting the \mstar~\msun\ subgiant,
\star. The properties of the host star are presented in \S~\ref{properties}, 
and we describe our observations and orbital solution in
\S~\ref{orbit}. We present the results of our photometric monitoring in
\S~\ref{photometry} and conclude with a discussion in \S~\ref{discussion}. 

\section{Sample}
\label{sample}

We have selected 159 HG stars based on the criteria
$0.5 < M_V < 3.5$, $0.55 < B-V < 1.0$, and $V \lesssim 7.6$, as listed in
the Hipparcos Catalog \citep{hipp}. Additionally, we selected only
stars lying more than 1~mag above the mean Hipparcos main--sequence, as
defined by \citet{wright04}. We chose the red cutoff to avoid
red giants, which are known to exhibit excess velocity jitter
\citep{frink01,hekker06}. The lower $M_V$ restriction 
avoids Cepheid variables, and the upper limit
excludes stars with masses less than $1.2$~\msun. We avoided stars in
the clump region of the H--R diagram ($B-V > 0.8$ and $M_V < 2.0$) in
order to avoid the mass ambiguity stemming from the closely--spaced,
overlapping isochrones between low--mass horizontal 
branch stars and high--mass stars on their first ascent to the giant
branch. 

\begin{figure}[t!]
\epsscale{1}
\plotone{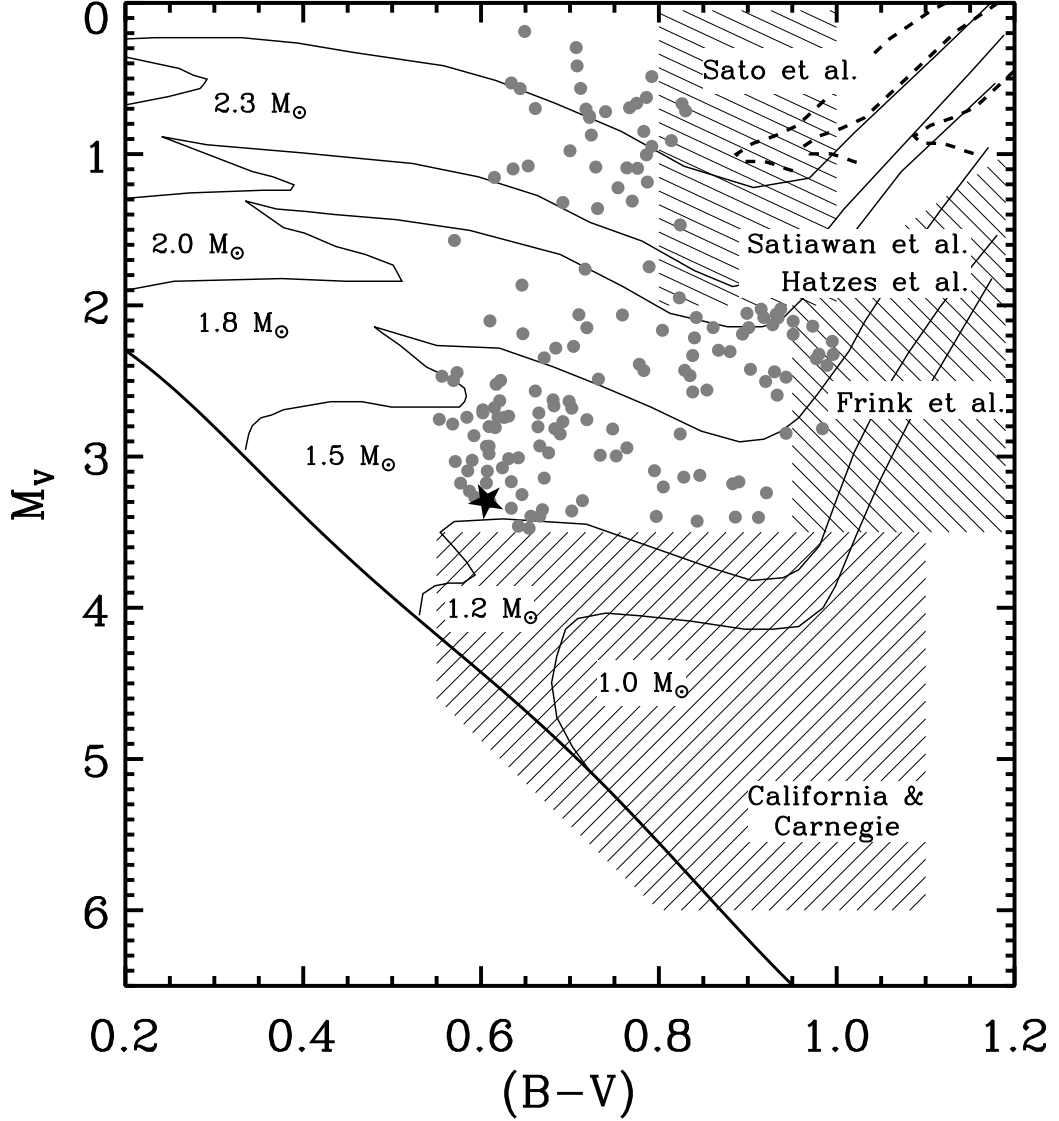}
\figcaption{Our sample of Hertzsprung Gap stars (filled circles) with
  respect to the mean Hipparcos main--sequence (thick line) and the
  theoretical mass tracks of \citet[][]{girardi02} for $\rm [Fe/H] = 0.0$. The
  mass tracks for Solar--mass stars on the horizontal branch with
  $\rm [Fe/H] = \{-0.7, -0.4, 0.0\}$ (from left to right) are shown as
  dashed lines in the upper right portion of the figure. The
  overlapping mass tracks in this region of the H--R diagram leads to
  ambiguious mass estimates. The   filled star shows 
  the position of \star. The hashed regions
  show the approximate ranges of the samples of other planet search
  programs. \label{sg_hr}}  
\end{figure}
\clearpage

The H--R diagram of our full sample is illustrated in
Figure~\ref{sg_hr}, along with the mass tracks of
\citet{girardi02}. The mass range of our sample is $1.2 < M_*
\lesssim 2.5$\msun, with a mean of 1.5~\msun. We are observing 115 of
the brightest and northernmost of these stars at Lick Observatory and the
remaining 44 at Keck Observatory.  Figure~\ref{sg_hr} also shows the
approximate search domains of 
other programs containing evolved, intermediate--mass stars. These
surveys include searches for planets around clump giants
\citep{sato03, setiawan03}, red giants \citep{frink02, hatzes05} and the
subgiants included in the CCPS. As evidenced from Figure~\ref{sg_hr},
HG stars occupy a unique and unexplored region in the H--R diagram.

\section{Properties of \star}
\label{properties}

\star\ (=HIP~96507) is a G0IV subgiant with $V=\vmag$, $B-V = \bv$, a
parallax--based distance of 47.6~pc, and an absolute 
magnitude $M_V = \mv$ \citep{hipp}. Its position in the H--R diagram in
relation to other stars in our sample is shown in Figure~\ref{sg_hr}. \star\
lies 1.1~mag above the mean Hipparcos main--sequence of stars in the
solar neighborhood \citep{wright04}, confirming its subgiant classification.
This star is chromospherically--quiet with $S = \sval$ and
$R^\prime_{HK} = \rhk$, from which we estimate a \prot~d 
rotation period. We used the LTE spectral synthesis
described by \citet{valenti05} to calculate 
$T_{eff} = \teff$~K,
$\rm [Fe/H] = \fe$,
$\log{g} = \logg$ and
$V_{rot}\sin{i} = \vsini$~\ks. We interpolated the star's
color, absolute 
magnitude and metallicity onto the stellar model grids of
\citet{girardi02} using the Bayesian methodology detailed by
\citet{pont04b}. Our interpolation yields an estimated stellar mass
$M_* = \mstar$\plmn0.1~\msun, radius $R_* = \rstar$~\plmn0.07~\rsun;
in agreement with $M_* = 1.31$~\msun\ and $R_* = 1.74$~\rsun\
estimated by \citet{allende99}. All of the stellar properties are
summarized in Table \ref{star_table}. 

\section{Observations and Orbit}
\label{orbit}

\begin{figure}[t!]
\epsscale{1}
\plotone{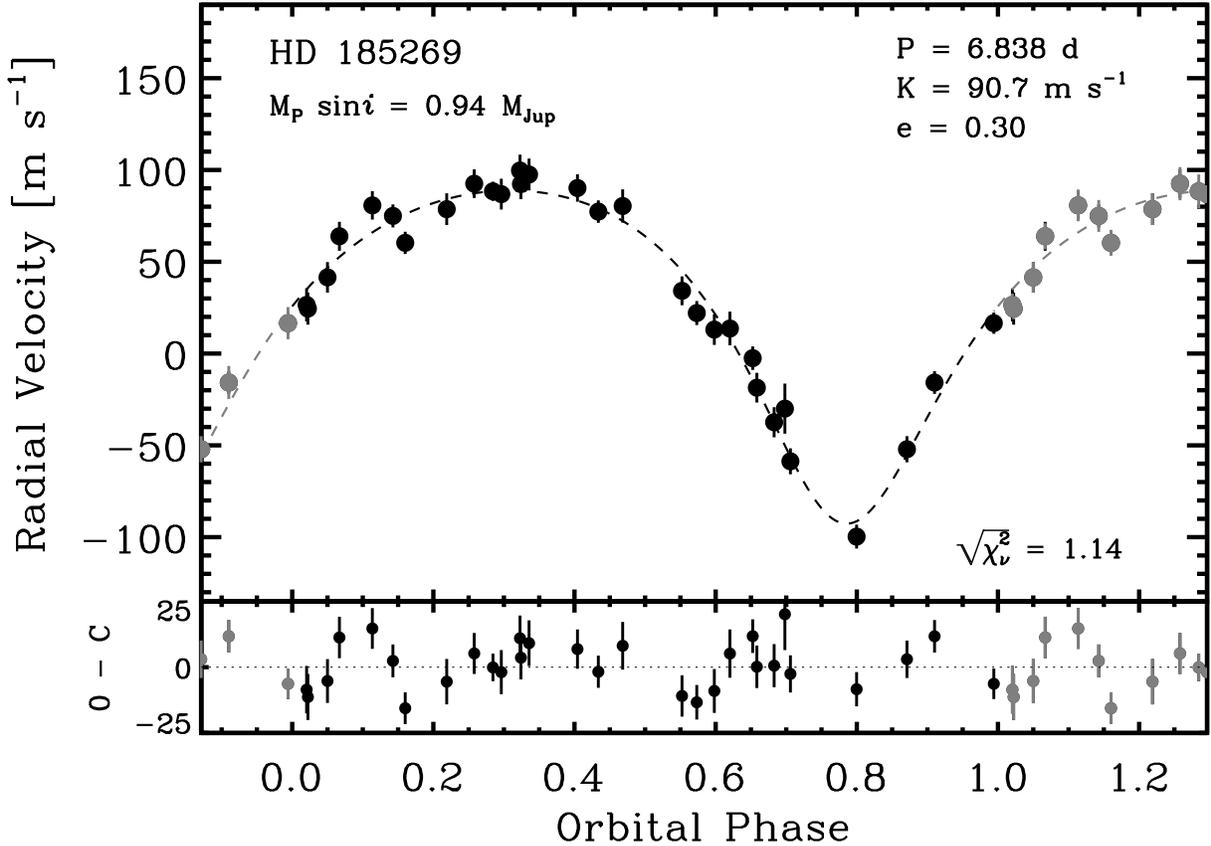}
\figcaption{Phased radial velocity measurements of \star. The dashed
  line shows the best--fit orbital solution. The bottom panel shows
  the phased residuals of the fit, which have RMS~$=$~\rms~\ms. \label{phased_orbit}} 
\end{figure}

We began observing \star\ in 2004 May at Lick Observatory using the
0.6~m Coude Auxiliary Telescope (CAT) and the 3~m Shane
telescope. Both telescopes feed the Hamilton echelle spectrograph,
thus no offset correction is needed for the velocities.
We measured radial velocities from the high--resolution
($\lambda/\delta\lambda = 50,000$) spectral observations using the 
\citet{butler96} iodine cell method. Traditionally, this method
requires an additional reference observation, or template, made
without the iodine cell. These template 
observations require higher signal and resolution than
normal radial velocity observations, which leads to increased exposure
times. Given our large target list and the small aperture of the CAT,
obtaining an observed template for 
each star would represent a prohibitive cost in observing
time. We instead analyze the initial 
observations of all our targets using synthetic (or ``morphed'')
templates, following the method described by \citet{johnson06}. Stars
showing excess RMS scatter are reanalyzed using a high--quality
observed template, obtained with the 3~m Shane telescope, to achieve
improved long--term velocity precision. 

The first five observations of \star\ spanned 1~yr and 
had an RMS scatter of 53~\ms, prompting us 
to initiate follow--up observations with an increased sampling rate
to search for a short--period signal. Our radial velocity
measurements are listed in Table~\ref{vel_table} along with the Julian
dates and internal measurement uncertainties. A periodogram analysis of
the velocities reveals a \period\ periodicity with a false alarm
probability $<0.001$\%.  
To search for a full orbital solution, we augmented our 
measurement uncertainties with a jitter estimate of \jitter~\ms, based
on the star's chromospheric activity index, absolute magnitude and color
\citep{saar98, wright05}. The best--fit Keplerian orbital solution
yields an orbital period $P = \period$~d, velocity semi--amplitude $K
= \kamp$~\ms\ and eccentricity $e = \ecc$\plmn\eerr. Using our stellar mass 
estimate of \mstar~\msun, we derive $M_P\sin{i} = \msini$~\mjup\ and $a =
\a$~AU. Figure \ref{phased_orbit} shows our velocities and
orbital solution phased with the \period~d period. The full orbital
solution and parameter uncertainties are listed in Table~\ref{orbit_table}.

We estimate the parameter uncertainties using a
Monte Carlo method. For each of 100 trials the best--fit Keplerian
is subtracted from the measured velocities. The residuals are then
scrambled and added back to the original measurements, and a new set
of orbital parameters is obtained. The standard deviations of the
parameters derived from all 
trials are adopted as the 1$\sigma$ uncertainties listed in
Table~\ref{orbit_table}. 

\section{Photometric Observations}
\label{photometry}

\citet{queloz01} and \citet{paulson04} have shown that active regions such 
as spots and plages on the photospheres of solar-type stars can cause 
low-amplitude, radial velocity variations (jitter) by distorting the stellar
line profiles as the spots are carried across the stellar disk by rotation.  
If a large active region exists for several stellar rotations (not an 
unusual circumstance for stars younger and more active than the Sun), then 
periodic rotational modulation of the spectral line profiles can mimic the 
presence of a planetary companion.  Therefore, precision photometric 
measurements can be an important complement to Doppler observations.  For
radial velocity variations caused by surface magnetic activity, the star 
will exhibit low-level photometric variability \citep[e.g.,][]{henry95} on
the radial velocity period.  If the radial velocity variability is the result
of true reflex motion caused by a planetary companion in orbit around the 
star, the star in general will not show photometric variability on the radial
velocity period \citep[e.g.,][]{henry00a}.  Photometric observations of
planetary-candidate stars can also detect transits of planetary companions 
with inclinations near 90\arcdeg\ and so allow the determination of 
a planet's true mass, radius, density, and composition 
\citep[e.g.,][]{henry00b,sato05,bouchy05}.

\begin{figure}[t!]
\epsscale{0.85}
\plotone{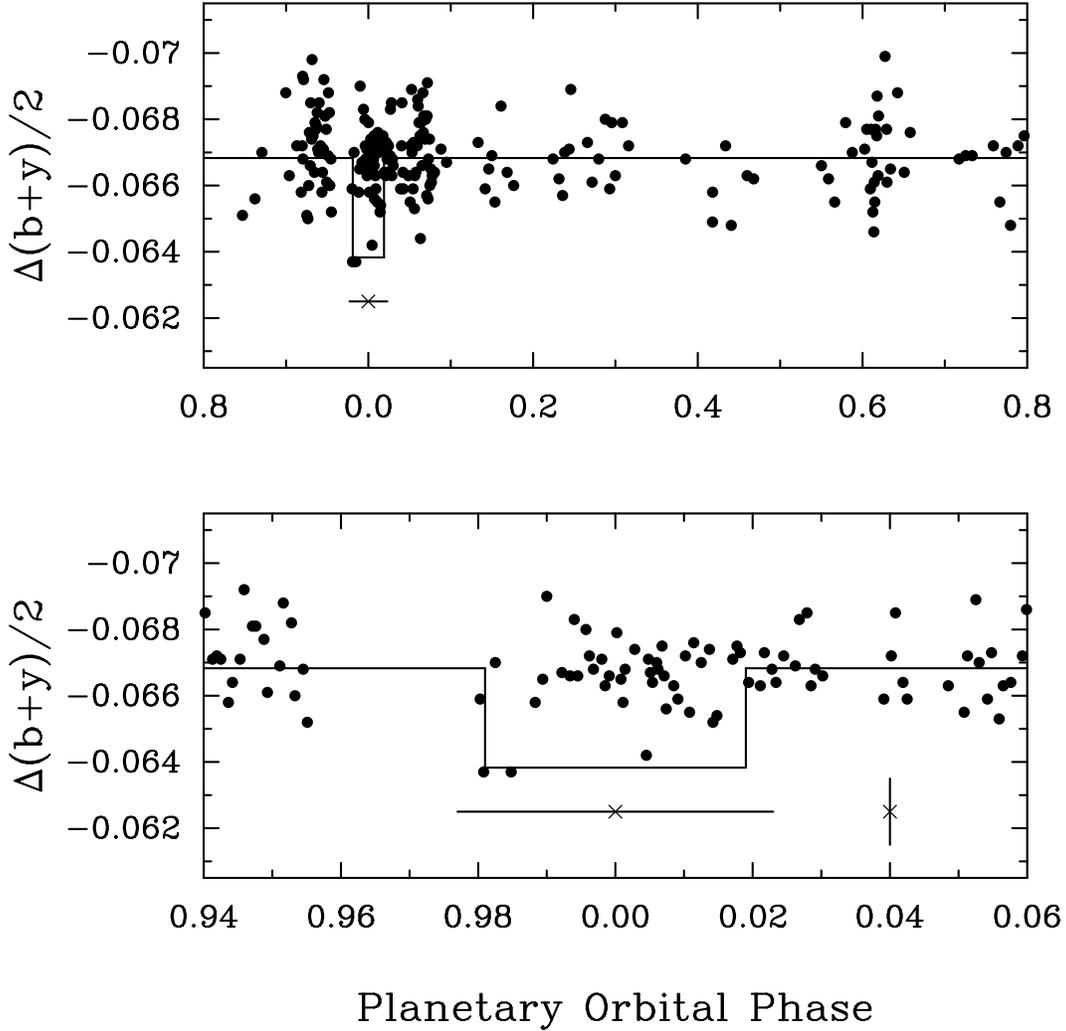}
\figcaption{Str\"omgren $(b+y)/2$ photometric observations of HD~185269
acquired with the T11 0.8~m APT at Fairborn Observatory.  The observations
have been phased to the orbital period of the planet and an estimated time 
of mid transit. ($Top$) There is no evidence for any periodicity in the 
observations between 1 and 100 days.  In particular, the star is constant 
on the radial velocity period to a limit of 0.0001 mag or so, supporting 
the planetary interpretation of the radial velocity variations.  
($Bottom$) The photometric observations around the predicted time of
transit replotted with an expanded scale on the abscissa.  The predicted 
transit depth of only 0.003 mag is shown for an assumed planetary radius 
of 1 R$_{Jup}$; such transits are ruled out by the observations. \label{phot_fig}}
\end{figure}

\clearpage

We observed \star\ with the T11 0.8~m automatic photometric telescope 
(APT) at Fairborn Observatory during May and June of 2006, obtaining a 
total of 207 brightness measurements.  The T11 APT is equipped 
with a two-channel precision photometer employing two EMI 9124QB bi-alkali 
photomultiplier tubes to make simultaneous measurements in the Str\"omgren 
$b$ and $y$ passbands.  The APT measures the difference in brightness between 
a program star and a nearby constant comparison star or stars with a 
typical precision of 0.0015 mag for bright stars ($V < 8.0$).  For \star, 
we used the two comparison stars HD\,184151 ($V$ = 6.87, $B-V$ = 0.46, F5 V)
and HD\,184381 ($V$ = 6.70, $B-V$ = 0.45, F5 V).  Differential magnitudes
between the two comparison stars showed them both to be constant to 0.0012 mag 
or better on a night-to-night timescale.  We created Str\"omgren $b$ and 
$y$ differential magnitudes of \star\ with respect to the {\it average} 
of the two comparison stars to improve our photometric precision.  The
differential magnitudes were reduced with nightly extinction coefficients 
and transformed to the Str\"omgren system with yearly mean transformation 
coefficients.  To improve precision still further, we combined the separate
$b$ and $y$ differential magnitudes into a single $(b+y)/2$ pass band.
Additional information on the telescope, photometer, observing procedures, 
and data reduction techniques employed with the T11 APT can be found in 
\citet{henry99} and \citet{eaton03}.

The 207 combined $(b+y)/2$ differential magnitudes of \star\ are plotted 
in the top panel of Figure~\ref{phot_fig}.  The observations are
phased with the planetary orbital period and the time of mid transit
given in Table~\ref{orbit_table}.  The standard deviation of the
observations from the mean brightness level is 0.0011 mag,  
suggesting that \star\ as well as its comparison stars are all highly 
constant.  Period analysis does not reveal any periodicity between
0.03 and 100  days.  A least-squares sine fit of the observations
phased to the radial  
velocity period gives a semi-amplitude of 0.00015 $\pm$ 0.00012 mag.  This 
very low limit to possible photometric variability supports planetary-reflex 
motion as the cause of the radial velocity variations.

In the bottom panel of Figure~\ref{phot_fig}, the observations near 
phase 0.0 are replotted with an expanded scale on the abscissa.  The solid
curve in each of the two  
panels approximates the predicted transit light curve assuming a planetary 
orbital inclination of 90\arcdeg\ (central transits).  The out-of-transit 
light level corresponds to the mean brightness of the observations.  The 
predicted transit duration is calculated from the orbital elements, while the 
predicted transit depth of 0.003 mag is derived from the stellar radius of 
1.88 R$_{\sun}$ from Table~\ref{star_table} and an assumed planetary radius 
of 1.0 R$_{Jup}$.
Thus, any transits of the planet across the subgiant star are expected to be
very shallow, as was the case with the transits of HD\,149026
\citep{sato05}.  The horizontal bar below the predicted transit window
represents the approximate uncertainty in the time of mid transit,
computed from the orbital elements.  The 
vertical error bar to the right of the transit window corresponds to the 
$\pm$ 0.0011 mag measurement uncertainties for a single observation.  The 
geometric probability of transits in this system is $\sim$12\%, computed 
from the orbital elements in Table~\ref{orbit_table} and equation 1 of
\citet{seagroves03}.  Although the uncertainty in the time of mid
transit is slightly larger than the predicted duration of any
transits, the observations nonetheless rule out the existence of
transits except perhaps for short events occurring around phase 0.97.
In the absence of transits, the orbital inclination must be less than
$\sim83\arcdeg$. 

\section{Summary and Discussion}
\label{discussion}

We are monitoring the radial velocities of a sample of 159
intermediate--mass Hertzsprung Gap (HG) stars in order to study the
relationship between stellar mass and planet occurrence rate.
We present the detection of a \msini~\mjup\ planet in an eccentric,
\period~d orbit around the \mstar~\msun\ subgiant \star. 

\begin{figure}[t!]
\epsscale{1}
\plotone{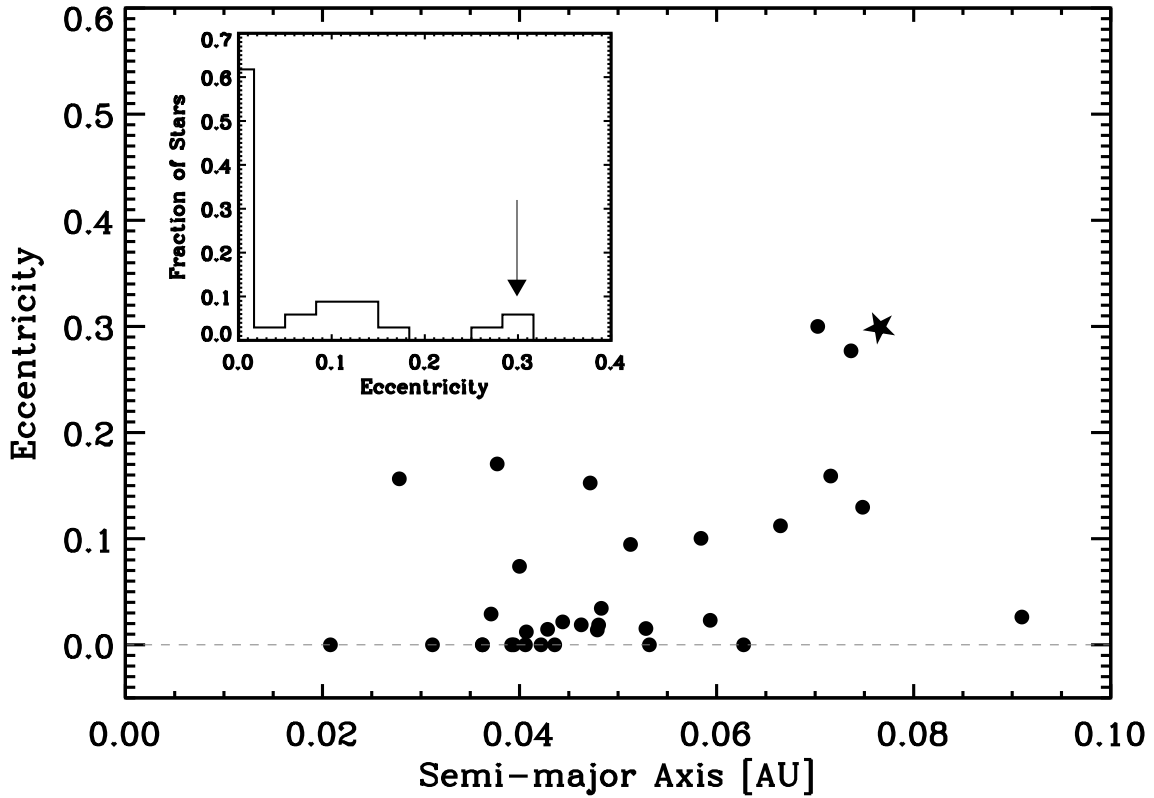}
\figcaption{The eccentricity--period distribution of hot jupiters
  listed in \citet{butler06}. \star\ is shown as a filled
  pentagram. The inset shows the histogram of eccentricities for the
  same sample of stars. The arrow denotes the position of \star. \label{ecc_plot}} 
\end{figure}

Compared to other planets with orbital separations $a<0.1$~AU (also
known as ``hot jupiters''), \star b has an unusually large
eccentricity. Figure~\ref{ecc_plot} shows the distribution
of eccentricities for the 33 planets having $a < 0.1$~AU
\citep{butler06}. With $e = \ecc$, \star\ b stands out from the
distribution as one of only three hot jupiters with eccentricities in
excess of 0.2. The other two planets are HD~162020\ b \citep[$e =
  0.277$;][]{udry02} and HD~118203\ b \citep[$e = 0.309$;][]{dasilva06}.

The origin of
eccentricities amongst single exoplanets is not well understood, but
the gravitational influence of additional companions can drive the
eccentricities of planets in multicomponent systems. 
An as--yet undetected second planet is likely the cause of the large
eccentricity of HD~118203b---\citet{dasilva06} report a linear trend
with a 49.7~\ms~yr$^{-1}$ slope. However, we do not 
see evidence of a linear trend in our observations of \star, nor
significant periodicities in the Keplerian fit residuals. We are
continuing to monitor this star at Keck and Lick Observatories to
search for additional low--mass companions. 

It is thought that the nearly circular orbits of most hot jupiters is
a result of tidal circularization. This leaves open the possibility
that the tidal circularization timescale at the orbital distance of
\star b is longer than the age of the star. Indeed, theoretical
predictions of the circularization timescale have 
a strong dependence with semi--major axis, typically $t_{circ} \propto
a^{13/3}$ \citep{terquem98}. Eight of the 10 planets with $a >
0.055$~AU have eccentricities greater than 0.1, which may indicate
that the timescale for tidal circularization 
at distances beyond $\sim 0.055$~AU is longer than the age of most 
FGK stars. The detection of additional hot jupiters by programs such as
the N2K Consortium \citep{fischer05a} will help shed light on the
relationship between the eccentricity and orbital separation of
short--period planets.

Other searches for planets around intermediate--mass stars 
have so far focused on the red giant branch \citep{frink02, hatzes05} and clump
regions of the H--R diagram \citep{sato03, setiawan03}. These programs
have to date discovered a total of 6 sub-stellar objects orbiting
giant stars, proving that 
planets do form and can be detected around evolved stars more massive than
$\sim 1.5$~\msun. However, stars in the clump and red giant branches
follow closely--spaced and often overlapping evolutionary
tracks, making precise stellar mass estimations from isochrone
interpolation difficult (see Figure~\ref{sg_hr}). By contrast, HG stars
have mass tracks that are nearly parallel and widely--spaced in $M_V$,
enabling precise mass determinations. HG stars also exhibit 
lower velocity jitter and have smaller radii than red giants
\citep{frink01, hekker06}, which enables the detection of a wider
range of planet masses and orbital separations. We will present the stellar 
characteristics and velocity behavior of our sample in a forthcoming
paper. 

\acknowledgements 

We extend our gratitude to the many CAT observers who have helped 
with this project, including Shannon Patel, Julia Kregenow, Karin
Sandstrom, Katie Peek and Bernie Walp. Special thanks to Conor Laver
and Franck Marchis for lending a portion of their 3m time to
observe this star before it set in 2005. We 
also gratefully acknowledge the efforts and dedication of the Lick
Observatory staff. We appreciate funding from NASA grant NNG05GK92G
(to GWM) for supporting this research. DAF is a Cottrell Science
Scholar of Research Corporation and acknowledges support from NASA
Grant NNG05G164G that made this work possible. GWH acknowledges
support from NASA grant NCC5-511 and NSF grant HRD-9706268.

\begin{deluxetable}{ll}
\tablecaption{Stellar Parameters\label{star_table}}
\tablewidth{0pt}
\tablehead{}
\startdata
V              & \vmag         \\
$M_V$          & \mv~(0.08)    \\
B-V            & \bv           \\
Spectral Type  & \sptype       \\
Distance (pc)  & \dist~(1.0)   \\
${\rm [Fe/H]}$ & \fe~(0.05)    \\
$T_{eff}$~(K)  & \teff~(50)   \\
$V_{rot}\sin{i}$~(\ks)     & \vsini~(0.5)     \\
$\log{g}$       & \logg~(0.07)    \\
$M_{*}$~(\msun) & \mstar~(0.1)   \\
$R_{*}$~(\rsun) & \rstar~(0.1)   \\
$S_{HK}$       & \sval             \\
$\log R'_{HK}$ & \rhk             \\
$P_{rot}$~(d)  & \prot         \\
Age~(Gyr)  & \age         \\
Jitter (\ms) & \jitter \\
\enddata
\end{deluxetable}

\begin{deluxetable}{lrr}
\tablecaption{Radial Velocities for HD~185269\label{vel_table}}
\tablewidth{0pt}
\tablehead{
\colhead{JD} &
\colhead{RV} &
\colhead{Uncertainty} \\
\colhead{-2440000} &
\colhead{(m~s$^{-1}$)} &
\colhead{(m~s$^{-1}$)} 
}
\startdata
13155.980 &    -6.5 &   5.8 \\
13211.839 &    30.7 &   6.2 \\
13522.875 &  -107.0 &   5.6 \\
13576.840 &   -34.9 &   7.2 \\
13619.734 &  -100.0 &   5.7 \\
13629.661 &    51.7 &   6.2 \\
13668.621 &   -21.5 &   6.7 \\
13718.641 &    49.6 &   6.3 \\
13720.587 &   -34.3 &   6.9 \\
13866.932 &   -23.5 &   6.4 \\
13867.878 &    11.8 &   4.3 \\
13868.806 &    38.9 &   5.9 \\
13868.996 &    44.3 &   5.6 \\
13869.982 &    32.4 &   6.8 \\
13879.841 &   -63.7 &   5.4 \\
13880.912 &    16.0 &   7.1 \\
13884.918 &   -50.7 &   5.6 \\
13885.924 &  -148.0 &   5.5 \\
13891.795 &   -66.5 &   6.9 \\
13891.963 &   -85.7 &   7.0 \\
13894.908 &    32.8 &   4.9 \\
13895.895 &    44.6 &   5.0 \\
13896.895 &    42.2 &   4.5 \\
13897.908 &   -13.8 &   5.1 \\
13898.905 &   -78.0 &  12.0 \\
13900.929 &   -31.9 &   5.2 \\
13901.947 &    26.7 &   5.7 \\
13902.915 &    40.5 &   4.6 \\
13903.936 &    29.1 &   5.6 \\
13904.888 &   -25.6 &   6.0 \\
\enddata
\end{deluxetable}

\begin{deluxetable}{lll}
\tablecaption{Orbital Parameters for \star\,b \label{orbit_table}}
\tablewidth{0pt}
\tablehead{
\colhead{} \\
}
\startdata
P~(d)               & \period~(\perr)       \\
T$_p$\tablenotemark{a}~(JD)          & \tp~(\terr)  \\
T$_{transit}$~(JD)          & \ttr~(\ttrerr)  \\
e                   & \ecc~(\eerr)      \\
K$_1$~(\ms)         & \kamp~(\kerr)      \\
$\omega$~(deg)      & \om~(\oerr)    \\
$M_P\sin{i}$~($M_{Jup}$)   & \msini           \\
$a$~(AU)             & \a            \\
Fit RMS~(\ms)       & \rms           \\
$\sqrt{\chi_\nu^2}$ & \chisq          \\
$N_{obs}$          & \nobseff \\
\enddata 
\tablenotetext{a}{Time of periastron passage.}
\end{deluxetable}

\begin{thebibliography}{50}
\expandafter\ifx\csname natexlab\endcsname\relax\def\natexlab#1{#1}\fi

\bibitem[{{Alibert} {et~al.}(2005){Alibert}, {Mordasini}, {Benz}, \&
  {Winisdoerffer}}]{alibert05}
{Alibert}, Y., {Mordasini}, C., {Benz}, W., \& {Winisdoerffer}, C. 2005, \aap,
  434, 343

\bibitem[{{Allende Prieto} \& {Lambert}(1999)}]{allende99}
{Allende Prieto}, C. \& {Lambert}, D.~L. 1999, \aap, 352, 555

\bibitem[{{Bouchy} {et~al.}(2005){Bouchy}, {Udry}, {Mayor}, {Moutou}, {Pont},
  {Iribarne}, {da Silva}, {Ilovaisky}, {Queloz}, {Santos}, {S{\'e}gransan}, \&
  {Zucker}}]{bouchy05}
{Bouchy}, F., {Udry}, S., {Mayor}, M., {Moutou}, C., {Pont}, F., {Iribarne},
  N., {da Silva}, R., {Ilovaisky}, S., {Queloz}, D., {Santos}, N.~C.,
  {S{\'e}gransan}, D., \& {Zucker}, S. 2005, \aap, 444, L15

\bibitem[{{Butler} {et~al.}(1996){Butler}, {Marcy}, {Williams}, {McCarthy},
  {Dosanjh}, \& {Vogt}}]{butler96}
{Butler}, R.~P., {Marcy}, G.~W., {Williams}, E., {McCarthy}, C., {Dosanjh}, P.,
  \& {Vogt}, S.~S. 1996, \pasp, 108, 500

\bibitem[{{Butler} {et~al.}(2006){Butler}, {Wright}, {Marcy}, {Fischer},
  {Vogt}, {Tinney}, {Jones}, {Carter}, {Johnson}, {McCarthy}, {Munoz}, \&
  {Penny}}]{butler06}
{Butler}, R.~P., {Wright}, J.~T., {Marcy}, G.~W., {Fischer}, D.~A., {Vogt},
  S.~S., {Tinney}, C.~G., {Jones}, H.~R.~A., {Carter}, B.~D., {Johnson}, J.~A.,
  {McCarthy}, C., {Munoz}, M., \& {Penny}, A.~J. 2006, \apj, submitted

\bibitem[{{da Silva} {et~al.}(2006){da Silva}, {Udry}, {Bouchy}, {Mayor},
  {Moutou}, {Pont}, {Queloz}, {Santos}, {S{\'e}gransan}, \&
  {Zucker}}]{dasilva06}
{da Silva}, R., {Udry}, S., {Bouchy}, F., {Mayor}, M., {Moutou}, C., {Pont},
  F., {Queloz}, D., {Santos}, N.~C., {S{\'e}gransan}, D., \& {Zucker}, S. 2006,
  \aap, 446, 717

\bibitem[{{do Nascimento} {et~al.}(2003){do Nascimento}, {Canto Martins},
  {Melo}, {Porto de Mello}, \& {De Medeiros}}]{donascimento03}
{do Nascimento}, J.~D., {Canto Martins}, B.~L., {Melo}, C.~H.~F., {Porto de
  Mello}, G., \& {De Medeiros}, J.~R. 2003, \aap, 405, 723

\bibitem[{{do Nascimento} {et~al.}(2000){do Nascimento}, {Charbonnel},
  {L{\`e}bre}, {de Laverny}, \& {De Medeiros}}]{donascimento00}
{do Nascimento}, J.~D., {Charbonnel}, C., {L{\`e}bre}, A., {de Laverny}, P., \&
  {De Medeiros}, J.~R. 2000, \aap, 357, 931

\bibitem[{{Eaton} {et~al.}(2003){Eaton}, {Henry}, \& {Fekel}}]{eaton03}
{Eaton}, J.~A., {Henry}, G.~W., \& {Fekel}, F.~C. 2003, The Future of Small
  Telescopes In The New Millennium.~Volume II - The Telescopes We Use, 189

\bibitem[{{Endl} {et~al.}(2006){Endl}, {Cochran}, {Kuerster}, {Paulson},
  {Wittenmyer}, {MacQueen}, \& {Tull}}]{endl06}
{Endl}, M., {Cochran}, W.~D., {Kuerster}, M., {Paulson}, D.~B., {Wittenmyer},
  R.~A., {MacQueen}, P.~J., \& {Tull}, R.~G. 2006, ArXiv Astrophysics e-prints

\bibitem[{{ESA}(1997)}]{hipp}
{ESA}, . 1997, VizieR Online Data Catalog, 1239, 0

\bibitem[{{Fischer} {et~al.}(2005){Fischer}, {Laughlin}, {Butler}, {Marcy},
  {Johnson}, {Henry}, {Valenti}, {Vogt}, {Ammons}, {Robinson}, {Spear},
  {Strader}, {Driscoll}, {Fuller}, {Johnson}, {Manrao}, {McCarthy},
  {Mu{\~n}oz}, {Tah}, {Wright}, {Ida}, {Sato}, {Toyota}, \&
  {Minniti}}]{fischer05a}
{Fischer}, D.~A., {Laughlin}, G., {Butler}, P., {Marcy}, G., {Johnson}, J.,
  {Henry}, G., {Valenti}, J., {Vogt}, S., {Ammons}, M., {Robinson}, S.,
  {Spear}, G., {Strader}, J., {Driscoll}, P., {Fuller}, A., {Johnson}, T.,
  {Manrao}, E., {McCarthy}, C., {Mu{\~n}oz}, M., {Tah}, K.~L., {Wright}, J.,
  {Ida}, S., {Sato}, B., {Toyota}, E., \& {Minniti}, D. 2005, \apj, 620, 481

\bibitem[{{Fischer} \& {Valenti}(2005)}]{fischer05b}
{Fischer}, D.~A. \& {Valenti}, J. 2005, \apj, 622, 1102

\bibitem[{{Frink} {et~al.}(2002){Frink}, {Mitchell}, {Quirrenbach}, {Fischer},
  {Marcy}, \& {Butler}}]{frink02}
{Frink}, S., {Mitchell}, D.~S., {Quirrenbach}, A., {Fischer}, D.~A., {Marcy},
  G.~W., \& {Butler}, R.~P. 2002, \apj, 576, 478

\bibitem[{{Frink} {et~al.}(2001){Frink}, {Quirrenbach}, {Fischer}, {R{\"o}ser},
  \& {Schilbach}}]{frink01}
{Frink}, S., {Quirrenbach}, A., {Fischer}, D., {R{\"o}ser}, S., \& {Schilbach},
  E. 2001, \pasp, 113, 173

\bibitem[{{Galland} {et~al.}(2005){Galland}, {Lagrange}, {Udry}, {Chelli},
  {Pepe}, {Queloz}, {Beuzit}, \& {Mayor}}]{galland05}
{Galland}, F., {Lagrange}, A.-M., {Udry}, S., {Chelli}, A., {Pepe}, F.,
  {Queloz}, D., {Beuzit}, J.-L., \& {Mayor}, M. 2005, \aap, 443, 337

\bibitem[{{Girardi} {et~al.}(2002){Girardi}, {Bertelli}, {Bressan}, {Chiosi},
  {Groenewegen}, {Marigo}, {Salasnich}, \& {Weiss}}]{girardi02}
{Girardi}, L., {Bertelli}, G., {Bressan}, A., {Chiosi}, C., {Groenewegen},
  M.~A.~T., {Marigo}, P., {Salasnich}, B., \& {Weiss}, A. 2002, \aap, 391, 195

\bibitem[{{Gonzalez}(1997)}]{gonzalez97}
{Gonzalez}, G. 1997, \mnras, 285, 403

\bibitem[{{Gray} \& {Nagar}(1985)}]{gray85}
{Gray}, D.~F. \& {Nagar}, P. 1985, \apj, 298, 756

\bibitem[{{Hatzes} {et~al.}(2005){Hatzes}, {Guenther}, {Endl}, {Cochran},
  {D{\"o}llinger}, \& {Bedalov}}]{hatzes05}
{Hatzes}, A.~P., {Guenther}, E.~W., {Endl}, M., {Cochran}, W.~D.,
  {D{\"o}llinger}, M.~P., \& {Bedalov}, A. 2005, \aap, 437, 743

\bibitem[{{Hekker} {et~al.}(2006){Hekker}, {Reffert}, {Quirrenbach},
  {Mitchell}, {Fischer}, {Marcy}, \& {Butler}}]{hekker06}
{Hekker}, S., {Reffert}, S., {Quirrenbach}, A., {Mitchell}, D.~S., {Fischer},
  D.~A., {Marcy}, G.~W., \& {Butler}, R.~P. 2006, ArXiv Astrophysics e-prints

\bibitem[{{Henry}(1999)}]{henry99}
{Henry}, G.~W. 1999, \pasp, 111, 845

\bibitem[{{Henry} {et~al.}(2000{\natexlab{a}}){Henry}, {Baliunas}, {Donahue},
  {Fekel}, \& {Soon}}]{henry00a}
{Henry}, G.~W., {Baliunas}, S.~L., {Donahue}, R.~A., {Fekel}, F.~C., \& {Soon},
  W. 2000{\natexlab{a}}, \apj, 531, 415

\bibitem[{{Henry} {et~al.}(1995){Henry}, {Fekel}, \& {Hall}}]{henry95}
{Henry}, G.~W., {Fekel}, F.~C., \& {Hall}, D.~S. 1995, \aj, 110, 2926

\bibitem[{{Henry} {et~al.}(2000{\natexlab{b}}){Henry}, {Marcy}, {Butler}, \&
  {Vogt}}]{henry00b}
{Henry}, G.~W., {Marcy}, G.~W., {Butler}, R.~P., \& {Vogt}, S.~S.
  2000{\natexlab{b}}, \apjl, 529, L41

\bibitem[{{Ida} \& {Lin}(2005{\natexlab{a}})}]{ida05a}
{Ida}, S. \& {Lin}, D.~N.~C. 2005{\natexlab{a}}, Progress of Theoretical
  Physics Supplement, 158, 68

\bibitem[{{Ida} \& {Lin}(2005{\natexlab{b}})}]{ida05b}
---. 2005{\natexlab{b}}, \apj, 626, 1045

\bibitem[{{Johnson} {et~al.}(2006){Johnson}, {Marcy}, {Fischer}, {Laughlin},
  {Butler}, {Henry}, {Valenti}, {Ford}, {Vogt}, \& {Wright}}]{johnson06}
{Johnson}, J.~A., {Marcy}, G.~W., {Fischer}, D.~A., {Laughlin}, G., {Butler},
  R.~P., {Henry}, G.~W., {Valenti}, J.~A., {Ford}, E.~B., {Vogt}, S.~S., \&
  {Wright}, J.~T. 2006, ArXiv Astrophysics e-prints

\bibitem[{{Kornet} {et~al.}(2006){Kornet}, {Wolf}, \& {Rozyczka}}]{kornet06}
{Kornet}, K., {Wolf}, S., \& {Rozyczka}, M. 2006, ArXiv Astrophysics e-prints

\bibitem[{{Laughlin} {et~al.}(2004){Laughlin}, {Bodenheimer}, \&
  {Adams}}]{laughlin04}
{Laughlin}, G., {Bodenheimer}, P., \& {Adams}, F.~C. 2004, \apjl, 612, L73

\bibitem[{{Laws} {et~al.}(2003){Laws}, {Gonzalez}, {Walker}, {Tyagi},
  {Dodsworth}, {Snider}, \& {Suntzeff}}]{laws03}
{Laws}, C., {Gonzalez}, G., {Walker}, K.~M., {Tyagi}, S., {Dodsworth}, J.,
  {Snider}, K., \& {Suntzeff}, N.~B. 2003, \aj, 125, 2664

\bibitem[{{Marcy} {et~al.}(2001){Marcy}, {Butler}, {Fischer}, {Vogt},
  {Lissauer}, \& {Rivera}}]{marcy01}
{Marcy}, G.~W., {Butler}, R.~P., {Fischer}, D., {Vogt}, S.~S., {Lissauer},
  J.~J., \& {Rivera}, E.~J. 2001, \apj, 556, 296

\bibitem[{{Paulson} {et~al.}(2004){Paulson}, {Saar}, {Cochran}, \&
  {Henry}}]{paulson04}
{Paulson}, D.~B., {Saar}, S.~H., {Cochran}, W.~D., \& {Henry}, G.~W. 2004, \aj,
  127, 1644

\bibitem[{{Pollack} {et~al.}(1996){Pollack}, {Hubickyj}, {Bodenheimer},
  {Lissauer}, {Podolak}, \& {Greenzweig}}]{pollack96}
{Pollack}, J.~B., {Hubickyj}, O., {Bodenheimer}, P., {Lissauer}, J.~J.,
  {Podolak}, M., \& {Greenzweig}, Y. 1996, Icarus, 124, 62

\bibitem[{{Pont} \& {Eyer}(2004)}]{pont04b}
{Pont}, F. \& {Eyer}, L. 2004, \mnras, 351, 487

\bibitem[{{Queloz} {et~al.}(2001){Queloz}, {Henry}, {Sivan}, {Baliunas},
  {Beuzit}, {Donahue}, {Mayor}, {Naef}, {Perrier}, \& {Udry}}]{queloz01}
{Queloz}, D., {Henry}, G.~W., {Sivan}, J.~P., {Baliunas}, S.~L., {Beuzit},
  J.~L., {Donahue}, R.~A., {Mayor}, M., {Naef}, D., {Perrier}, C., \& {Udry},
  S. 2001, \aap, 379, 279

\bibitem[{{Saar} {et~al.}(1998){Saar}, {Butler}, \& {Marcy}}]{saar98}
{Saar}, S.~H., {Butler}, R.~P., \& {Marcy}, G.~W. 1998, \apjl, 498, L153+

\bibitem[{{Santos} {et~al.}(2004){Santos}, {Israelian}, \& {Mayor}}]{santos04}
{Santos}, N.~C., {Israelian}, G., \& {Mayor}, M. 2004, \aap, 415, 1153

\bibitem[{{Sato} {et~al.}(2003){Sato}, {Ando}, {Kambe}, {Takeda}, {Izumiura},
  {Masuda}, {Watanabe}, {Noguchi}, {Wada}, {Okada}, {Koyano}, {Maehara},
  {Norimoto}, {Okada}, {Shimizu}, {Uraguchi}, {Yanagisawa}, \&
  {Yoshida}}]{sato03}
{Sato}, B., {Ando}, H., {Kambe}, E., {Takeda}, Y., {Izumiura}, H., {Masuda},
  S., {Watanabe}, E., {Noguchi}, K., {Wada}, S., {Okada}, N., {Koyano}, H.,
  {Maehara}, H., {Norimoto}, Y., {Okada}, T., {Shimizu}, Y., {Uraguchi}, F.,
  {Yanagisawa}, K., \& {Yoshida}, M. 2003, \apjl, 597, L157

\bibitem[{{Sato} {et~al.}(2005){Sato}, {Fischer}, {Henry}, {Laughlin},
  {Butler}, {Marcy}, {Vogt}, {Bodenheimer}, {Ida}, {Toyota}, {Wolf}, {Valenti},
  {Boyd}, {Johnson}, {Wright}, {Ammons}, {Robinson}, {Strader}, {McCarthy},
  {Tah}, \& {Minniti}}]{sato05}
{Sato}, B., {Fischer}, D.~A., {Henry}, G.~W., {Laughlin}, G., {Butler}, R.~P.,
  {Marcy}, G.~W., {Vogt}, S.~S., {Bodenheimer}, P., {Ida}, S., {Toyota}, E.,
  {Wolf}, A., {Valenti}, J.~A., {Boyd}, L.~J., {Johnson}, J.~A., {Wright},
  J.~T., {Ammons}, M., {Robinson}, S., {Strader}, J., {McCarthy}, C., {Tah},
  K.~L., \& {Minniti}, D. 2005, \apj, 633, 465

\bibitem[{{Schrijver} \& {Pols}(1993)}]{schrijver93}
{Schrijver}, C.~J. \& {Pols}, O.~R. 1993, \aap, 278, 51

\bibitem[{{Seagroves} {et~al.}(2003){Seagroves}, {Harker}, {Laughlin}, {Lacy},
  \& {Castellano}}]{seagroves03}
{Seagroves}, S., {Harker}, J., {Laughlin}, G., {Lacy}, J., \& {Castellano}, T.
  2003, \pasp, 115, 1355

\bibitem[{{Setiawan} {et~al.}(2003){Setiawan}, {Hatzes}, {von der L{\" u}he},
  {Pasquini}, {Naef}, {da Silva}, {Udry}, {Queloz}, \& {Girardi}}]{setiawan03}
{Setiawan}, J., {Hatzes}, A.~P., {von der L{\" u}he}, O., {Pasquini}, L.,
  {Naef}, D., {da Silva}, L., {Udry}, S., {Queloz}, D., \& {Girardi}, L. 2003,
  \aap, 398, L19

\bibitem[{{Terquem} {et~al.}(1998){Terquem}, {Papaloizou}, {Nelson}, \&
  {Lin}}]{terquem98}
{Terquem}, C., {Papaloizou}, J.~C.~B., {Nelson}, R.~P., \& {Lin}, D.~N.~C.
  1998, \apj, 502, 788

\bibitem[{{Udry} {et~al.}(2003){Udry}, {Mayor}, {Clausen}, {Freyhammer},
  {Helt}, {Lovis}, {Naef}, {Olsen}, {Pepe}, {Queloz}, \& {Santos}}]{udry03}
{Udry}, S., {Mayor}, M., {Clausen}, J.~V., {Freyhammer}, L.~M., {Helt}, B.~E.,
  {Lovis}, C., {Naef}, D., {Olsen}, E.~H., {Pepe}, F., {Queloz}, D., \&
  {Santos}, N.~C. 2003, \aap, 407, 679

\bibitem[{{Udry} {et~al.}(2002){Udry}, {Mayor}, {Naef}, {Pepe}, {Queloz},
  {Santos}, \& {Burnet}}]{udry02}
{Udry}, S., {Mayor}, M., {Naef}, D., {Pepe}, F., {Queloz}, D., {Santos}, N.~C.,
  \& {Burnet}, M. 2002, \aap, 390, 267

\bibitem[{{Valenti} \& {Fischer}(2005)}]{valenti05}
{Valenti}, J.~A. \& {Fischer}, D.~A. 2005, \apjs, 159, 141

\bibitem[{{Wright}(2004)}]{wright04}
{Wright}, J.~T. 2004, \aj, 128, 1273

\bibitem[{{Wright}(2005)}]{wright05}
---. 2005, \pasp, 117, 657

\bibitem[{{Wright} {et~al.}(2004){Wright}, {Marcy}, {Butler}, \&
  {Vogt}}]{wright04b}
{Wright}, J.~T., {Marcy}, G.~W., {Butler}, R.~P., \& {Vogt}, S.~S. 2004, \apjs,
  152, 261

\end{thebibliography}
\end{document}